\documentclass{emulateapj}
\usepackage{epsfig}
\def\lap{\lower.5ex\hbox{$\; \buildrel < \over \sim \;$}}
\def\gap{\lower.5ex\hbox{$\; \buildrel > \over \sim \;$}}

\begin{document}

\title{The ACS Nearby Galaxy Survey Treasury VII. The NGC 4214 Starburst and the Effects of Star Formation History on Dwarf Morphology}

\author{Benjamin F. Williams\altaffilmark{1},
Julianne J. Dalcanton\altaffilmark{1},
Karoline M. Gilbert\altaffilmark{1},
Anil C. Seth\altaffilmark{2},
Daniel R. Weisz\altaffilmark{1},
Evan D. Skillman\altaffilmark{3},
Andrew E. Dolphin\altaffilmark{4}
}
\altaffiltext{1}{Department of Astronomy, Box 351580, University of Washington, Seattle, WA 98195; ben@astro.washington.edu; jd@astro.washington.edu; kgilbert@astro.washington.edu; dweisz@astro.washington.edu}
\altaffiltext{2}{CfA Fellow, Harvard-Smithsonian Center for Astrophysics, 60 Garden Street, Cambridge, MA 02138; aseth@cfa.harvard.edu}
\altaffiltext{3}{Department of Astronomy, University of Minnesota, 116 Church
St. SE, Minneapolis, MN 55455; skillman@astro.umn.edu}
\altaffiltext{4}{Raytheon, 1151 E. Hermans Road, Tucson, AZ 85756; adolphin@raytheon.com}


\begin{abstract}

We present deep Hubble Space Telescope Wide Field Planetary Camera 2 (WFPC2) optical observations obtained as part of the ACS Nearby Galaxy Survey Treasury (ANGST) as well as early release Wide Field Camera 3 (WFC3) ultra-violet and infrared observations of the nearby dwarf starbursting galaxy NGC~4214.  Our data provide a detailed example of how covering such a broad range in wavelength provides a powerful tool for constraining the physical properties of stellar populations.  The deepest data reach the ancient red clump at M$_{F814W}\sim$-0.2.  All of the optical data reach the main sequence turnoff for stars younger than $\sim$300~Myr, and the blue He burning sequence for stars younger than 500 Myr. The full CMD-fitting analysis shows that all three fields in our data set are consistent with $\sim$75\% of the stellar mass being older than 8~Gyr, in spite of showing a wide range in star formation rates at the present day.  Thus, our results suggest that the scale length of NGC~4214 has remained relatively constant for many Gyr.  As previously noted by others, we also find the galaxy has recently ramped up production, consistent with its bright UV luminosity and its population of UV-bright massive stars.  In the central field we find UV point sources with F336W magnitudes as bright as -9.9.  These are as bright as stars with masses of at least 52--56~M$_{\odot}$ and ages near 4~Myr in stellar evolution models. Assuming a standard IMF, our CMD is well-fitted by an increase in star formation rate beginning 100~Myr ago.  The stellar populations of this late-type dwarf are compared with those of NGC~404, an early-type dwarf that is also the most massive galaxy in its local environment.  The late-type dwarf appears to have a similar high fraction of ancient stars, suggesting that these dominant galaxies may form at early epochs even if they have low total mass and very different present-day morphologies.

\end{abstract}

\section{Introduction}

Irregular galaxies are the dominant type of star forming galaxy by
number.  Despite their typically isolated environments and relatively
low mass, the star formation histories (SFHs) of irregular galaxies
are surprisingly complex and varied.  In the context of hierarchical
structure formation, the formation epoch of these galaxies is expected
to be younger than that of more massive galaxies, a phenomenon
referred to as cosmic downsizing.  Deep resolved stellar photometry
allows us to measure their formation epoch, and to examine their
recent SFHs in detail.


Historically, early-type dwarfs were thought to be dominated by
ancient populations \citep{baade1963,zinn1980}, while dIrrs have been
considered to be the result of continuous, relatively constant, star
formation \citep[e.g.][]{hunter1982,hunter1997}.  This picture has
broken down in recent decades.  Early-type dwarfs were found to harbor
significant populations of intermediate-age stars \citep[see][for a
review, and many other studies]{mateo1998}, and recent studies of
dIrrs strongly suggest that they undergo episodic star formation
\citep[e.g.][and references therein]{stinson2007}.

If dIrrs are characterized by relatively constant star formation, we
might expect the fraction of ancient stars to be significantly lower
in dIrrs compared to early-type dwarfs.  However, resolved stellar
photometry of local dwarfs has shown that the fraction of ancient
stars is roughly the same in both morphological types
\citep{weisz2010}.  Other resolved stellar population studies suggest
bursts of about 500 Myr duration play an important role in the
evolution of dwarfs \citep{mcquinn2010a,mcquinn2010b}. It is therefore
possible that the presence of recent star formation episodes may be
the {\it only} difference between these classes of dwarf galaxies and
that such recent episodes of star formation actually contribute very
little to the total stellar mass of the systems.  Therefore, the
difference between these distinct morphological types may be due to
only a very small fraction of the stellar mass.  To further
investigate the possibility that the vast majority of the stellar mass
in dIrrs can be as old as the stellar mass of early type dwarfs and
that the star formation events witnessed in dIrrs are episodic, we
focus our attention on the nearby galaxy NGC~4214.

NGC~4214 is classified as an irregular galaxy
\citep[IAB(s)m,][]{devaucouleurs1991} with a mass of
$\sim$1.5$\times$10$^9$ M$_{\odot}$ \citep{karachentsev2004}, similar
to the Large Magellanic Cloud.  It is one of the nearest examples of a
starburst, and was the host of a Type~I supernova \citep{wellmann1955}
and recent nova \citep{humphreys2010}. It therefore has been studied
in detail, especially at UV and NIR wavelengths that probe the young
stellar populations.  For example, \citet{fanelli1997} were able to
make use of far-UV imaging from the Ultraviolet Imaging Telescope and
ground-based I-band to decompose the galaxy into a centrally
concentrated intense star forming component and an extended smooth
disk.  They suggest the two component structure is the result of a
recent merger. 

The bright resolved stellar populations have also been characterized.
A dominant red giant branch was detected, and a large metallicity
spread was measured by \citet{drozdovsky2002} using WFPC2 and NICMOS
photometry. \citet{ubeda2007a,ubeda2007b} used HST/WFPC2 and HST/STIS
imaging to suggest that the stellar initial mass function appears to
be steeper than that of \citet{salpeter1955} at high masses
($>$20~M$_{\odot}$), even though the galaxy's metallicity is similar
to that of the Small Magellanic Cloud.  However, it is unclear how
this slope is affected by fluctuations in the star formation rate.

We further examine the stellar populations of NGC~4214 using deeper
data than previous studies.  The new data include two HST WFPC2 fields
of the outer disk and one WFC3 field of the inner regions.  Through
detailed stellar evolution model fitting, we measure the star
formation history (SFH) of these regions and look for differences
between quiescent and intensely star forming regions.  Section 2
describes our data set and photometry.  Section 3 presents the results
of our model fitting procedure.  Section 4 discusses the results of
the measurements in the context of galaxy morphology and environment,
and Section 5 summarizes our conclusions.  We assume $(m-M)_0 =$
3.03~Mpc \citep{dalcanton2009} for conversions of angular measurements
to physical distances and assume an inclination of 38$^{\circ}$
\citep{karachentsev2004} for surface density measurements.  We adopt a
five-year WMAP \citep{dunkley2009} cosmology for all conversions
between time and redshift.

\section{Data}\label{data}

\begin{figure}
\centerline{\psfig{file=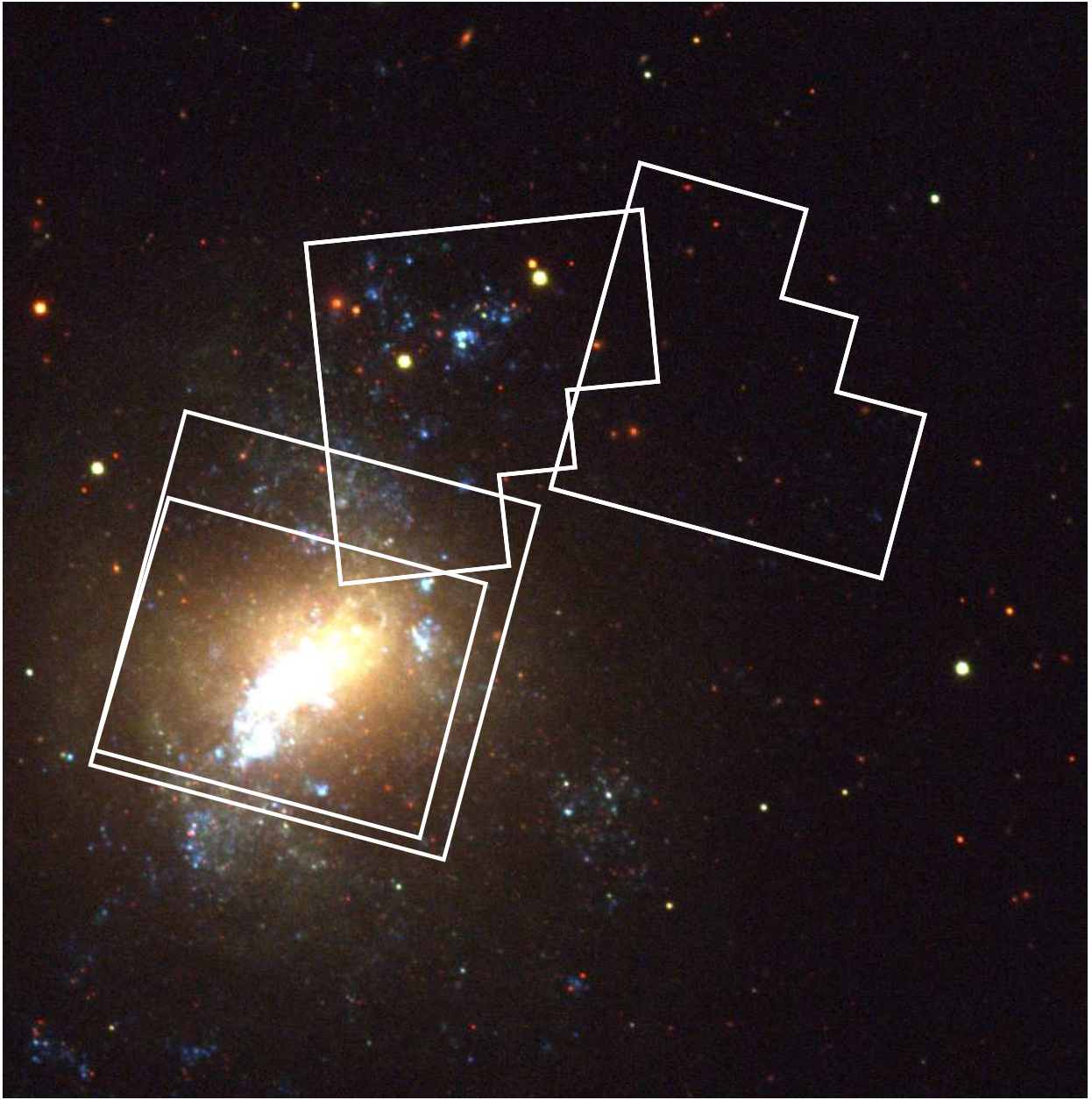,width=6.5in,angle=0}}
\caption{The locations of our NGC~4214 fields are shown on a 3-color
image using Sloan Digital Sky Survey $u'$ (blue), $g'$ (green), and
$i'$ (red) images. North is up.  East is left.  The small central
field is WFC3/IR.  The larger central field is WFC3/UVIS.  The outer
fields are WFPC2, with the westernmost being the deepest.}
\label{field_loc}
\end{figure}

As part of the ANGST program after the failure of ACS, from
2007-Dec-04 to 2007-Dec-23, we performed deep WFPC2 observations a
field in the NGC~4214 disk located at R.A.~(2000)=183.847375
(12:15:23.4), decl.~(2000)=36.362333 (+36:21:44) with a rotation angle
PA\_V3=119.9 degrees (GO-10915).  To improve our radial coverage to be
closer to the goal of the original ANGST program, from 2009-Feb-23 to
2009-Feb-26, we performed shallower supplemental observations for a
field located at R.A.~(2000)=183.878103 (12:15:30.7),
decl.~(2000)=36.356808 (+36:21:24.5) with a rotation angle PA\_V3=51.0
(GO-11986).  At the same time, WFC3/IR and WFC3/UVIS imaging of the
central area of the galaxy, proposed by the Scientific Oversight
Committee (SOC), were released.  Both fields were observed from
2009-Dec-22 to 2009-Dec-23 located at R.A.~(2000) =183.913333
(12:15:39.2), decl.~(2000) = 36.326944 (+36:19:37.0) with a rotation
angle PA\_V3=120.0 (GO-11360, PI: O'Connell).  Figure~\ref{field_loc}
shows outlines of the fields' locations.  Our field locations were
chosen to maximize the number of disk stars and avoid crowding.

In the deep field, we obtained 15 full-orbit exposures with the WFPC2
through the F606W (wide $V$) filter, and 29 full-orbit exposures
through the F814W ($I$ equivalent) filter.  These data totaled 39000~s
and 75400~s of exposure time in F606W and F814W, respectively.  In the
other WFPC2 field, we obtained 2 orbits through F606W, totaling
4800~s, and 4 orbits through F814W, totaling 9600~s. The WFC3/IR data
contained 1198~s and 2398~s of exposure in F110W and F160W,
respectively.  The WFC3/UVIS data contained 1683~s, 1540~s, and 1339~s
in F336W, F438W, and F814W, respectively.  All WFPC2 images were
calibrated in the HST pipeline with CALWP2 using OPUS version 2006\_6a
for the 2007 data and 2008\_5c for the 2009 data. All WFC3 images were
calibrated in the HST pipeline with CALWF3 version 2.0.

All of the data were processed through the ACS Nearby Galaxy Survey
Treasury (ANGST) data analysis pipeline \citep{dalcanton2009}, updated
to include WFC3 UVIS and IR data \citep{dalcanton2011}. As a brief
summary, the photometry was measured simultaneously for all of the
objects in the uncombined images using the software packages HSTPHOT
(for WFPC2) and DOLPHOT~1.2 \citep{dolphin2000,dolphin2009} including
the WFC3 module.  These packages are optimized for measuring
photometry of stars on HST images using the well-characterized and
stable point spread function (PSF) calculated with
TinyTim.\footnote{http://www.stsci.edu/software/tinytim/} The software
fits the PSF to all of the stars in each individual frame to find PSF
magnitudes.  It then determines and applies the aperture correction
for each image using the most isolated stars, corrects for the charge
transfer efficiency of the WFPC2 detector\footnote{July, 2008
formulae\\
http://purcell.as.arizona.edu/wfpc2\_calib/2008\_07\_19.html},
combines the results from the individual exposures, and converts the
measured count rates to the VEGAmag system.  

Our photometric error and completeness were then assessed by running
at least 10$^6$ artificial star tests for each field, in which a star
of known color and magnitude was placed into the images and the
photometry rerun to determine if the star was recovered and, if so,
the difference between the input and output magnitude.

The photometry output was then filtered to only allow objects
classified as stars with signal-to-noise $>$4 in both filters.  The
list was further culled using sharpness ($|F606W_{sharp} +
F814W_{sharp}| < 0.27$ for WFPC2, $|F336W_{sharp} + F438W_{sharp}| <
0.27$ for UVIS, and $|F110W_{sharp} + F160W_{sharp}| < 0.35$ for IR)
and crowding ($F606W_{crowd} + F814W_{crowd} < 0.7$ for WFPC2,
$F336W_{crowd} + F438W_{crowd} < 0.7$ for UVIS, and $F110W_{crowd} +
F160W_{crowd} < 0.48$ for IR).  The sharpness parameter returns zero
if a star is perfectly fit, negative if it is broader, and positive if
it is sharper than a typical star \citep{dolphin2000}. The crowding
parameter gives the difference between the magnitude of a star
measured before and after subtracting the neighboring stars in the
image.  When this value is large, it suggests that the star's
photometry was significantly affected by crowding, and we therefore
exclude it from our catalog.  The values for the sharpness and
crowding cuts for WFPC2 were the standard used by the ANGST program.
The cuts for WFC3 were chosen by looking at CMDs resulting from
different possibilities and choosing values that produced clean CMD
features and a low number of contaminants, falling outside of any
known feature. We note that our tests of different cuts showed that
the choices of cuts have little impact on our final SFH measurements
outside of the most recent time bin ($<$10 Myr), most likely due to
the significant clustering of the very youngest stars.  Our final
optical star catalogs contained 16806, 28088, 138350, and stars for
the outer (WFPC2 N), bridge (WFPC2 NW), and central (UVIS) fields,
respectively.

\begin{figure}
\centerline{\psfig{file=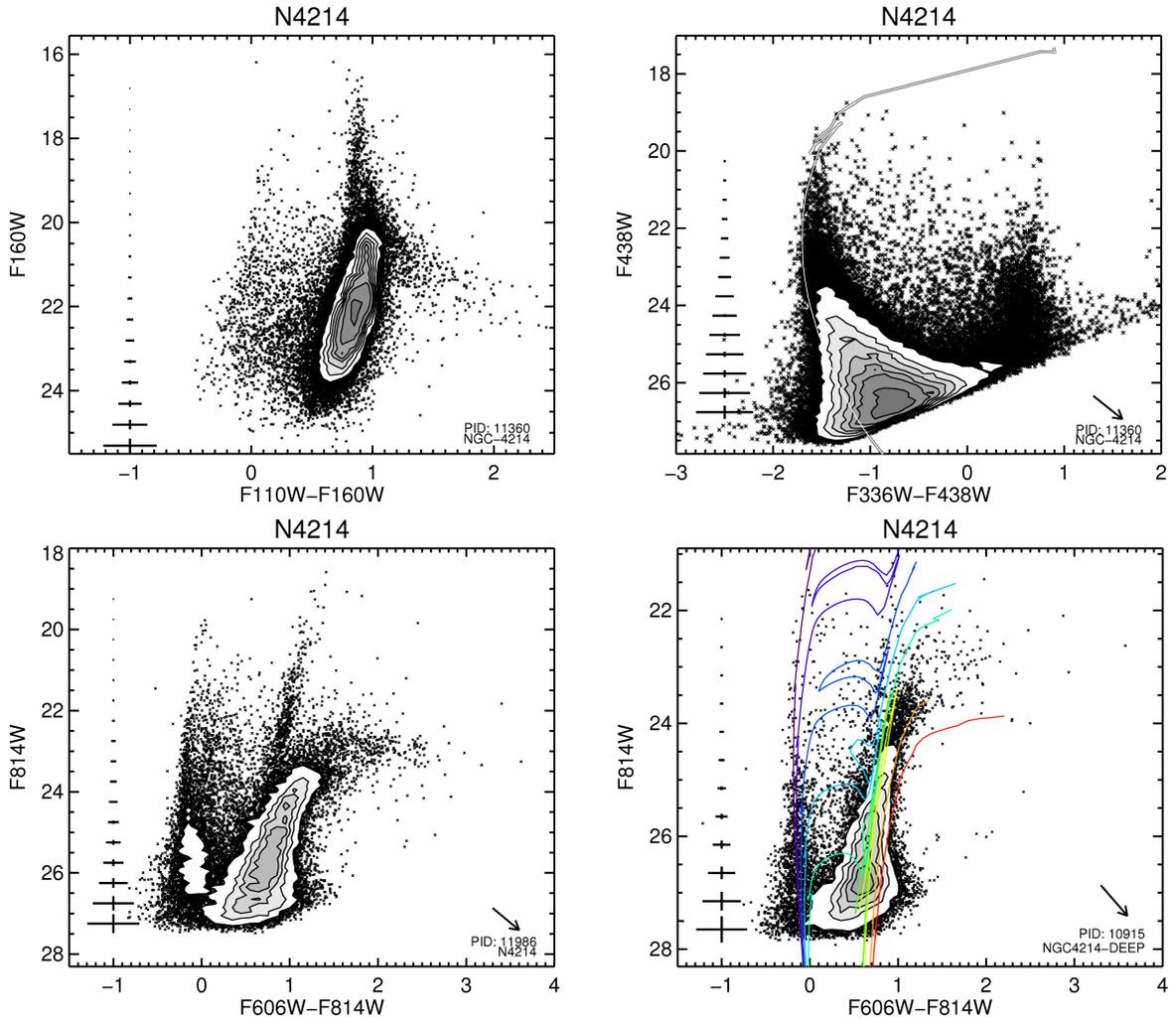,width=6.5in,angle=0}}
\caption{The CMDs of our 4 fields (\S~\ref{data}).  Top Left: WFC3/IR.
Top Right: WFC3/UVIS. A single isochrone for an age of 4~Myr and a
metallicity of [M/H]=-0.4 is overplotted (\S~\ref{bright}).  Arrow
indicates direction and magnitude of $E_{U-B}$=0.3.  Bottom Left: the
northern WFPC2 field. Arrow indicates direction and magnitude of
$E_{F606W-F814W}$=0.3. Bottom Right: the northwestern WFPC2 field with
isochrones overplotted (from blue to red: [M/H]=-0.4 and
log(age)~=~7.3,7.6,8.0,8.3,8.6, followed by log(age)=10.0 and
[M/H]~=~-1.7,-1.3,-0.7,-0.4, respectively). Arrow indicates direction
and magnitude of $E_{F606W-F814W}$=0.3.}
\label{cmds}
\end{figure}

Our final UV catalog for the central field contained 77656 stars, and
our final IR catalog for the central field contained 19807 stars.  In
the end, our analysis of the optical (F438W and F814W) photometry from
UVIS yielded results that were consistent with those of the UV (F336W
and F438W) photometry from UVIS.  Since no additional insight was
gained from the optical photometry, we will not discuss it further but
include the SFH and limiting magnitudes for reference.  Any statements
made about the UV photometry from UVIS was confirmed with the optical
photometry from UVIS.  The final CMDs are shown in Figure~\ref{cmds}.

\subsection{Field Division}

\begin{figure}
\centerline{\psfig{file=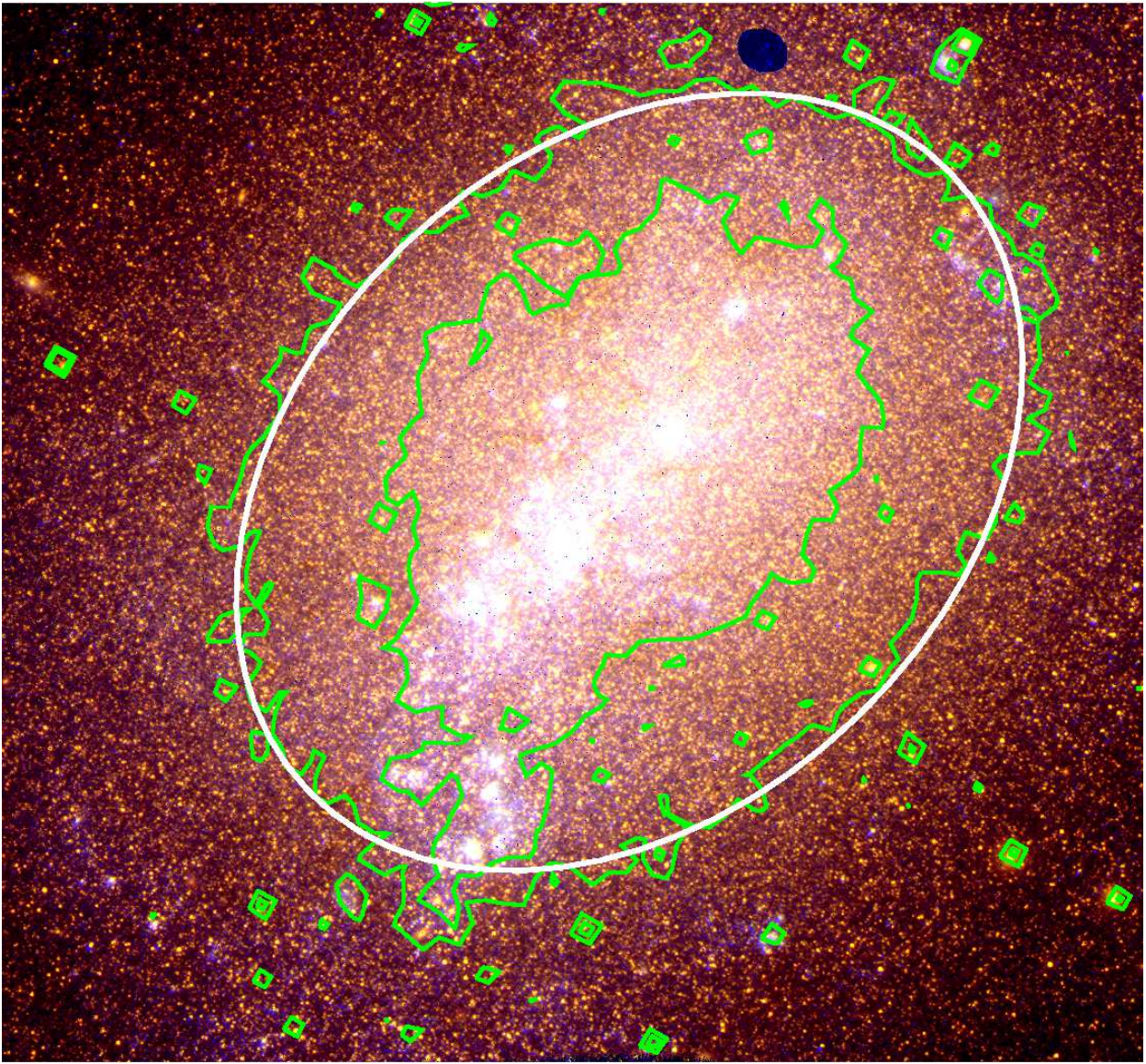,width=6.5in,angle=0}}
\caption{A WFC3 image of the central 2$'$X2$'$ of NGC~4214 shown in a
3-color image using WFC3 $F438W$ (blue), $F110W$ (green), and $F160W$
(red) images. North is up.  East is left.  Isophotal contours are
shown in green. Our field division ellipse is shown in white.  The
inner region is much more crowded and dusty than the outer region,
making CMD fitting more reliable when the two regions are modeled
separately.}
\label{ellipses}
\end{figure}

While all the WFPC2 imaging of the outer portions of the disk provided
high-quality photometry going much fainter than the tip of the red
giant branch (and even fainter than the red clump in the deep field),
the central WFC3 fields were not as consistent.  The UVIS images
proved to have similar completeness limits throughout the field, but
the IR images were clearly over-crowded near the galaxy center.  This
strong differential crowding in the IR data made analysis of the full
IR field dubious. We therefore separated the innermost portion of the
central field into two spatial components, following the isophotes of
the galaxy.  Our division ellipse and the galaxy isophotes in the
F110W image are shown in Figure~\ref{ellipses}.


\section{CMD Fitting and Temporal Resolution}

We measured the star formation rate and metallicity as a function of
stellar age using the software package MATCH \citep{dolphin2002} to
fit the observed CMDs.  We adopted magnitude cuts set to the limits
provided in Table~\ref{table}, and then fitted the stellar evolution
models of \citet[][with updates in \citealp{marigo2008} and
\citealp{girardi2008}]{girardi2002}, convolved with our photometric
error and completeness statistics and populated with a
\citet{salpeter1955} initial mass function (IMF).  The choices of
software and models used for the ANGST project are discussed in detail
in \citet{williams2009a}.  Briefly, comparisons of results when
fitting CMDs using different models and fitting software have shown
that the results are consistent within the estimated uncertainties;
furthermore, our choice of models provides the largest range of ages
and metallicities publicly available.

We first fit the data assuming a single foreground reddening
$A_V$=0.07 and distance $m-M_0$=27.41, which were obtained from the
ANGST survey \citep{dalcanton2009}, and the \citet{schlegel1998}
Galactic dust maps.  The best fit provides the relative contribution
of stars of each age and metallicity in each field. For the shallowest
data, the IR and optical data for the innermost region, we limited the
number of free parameters by imposing an ``increasing metallicity''
constraint on the fit, whereby the metallicity of the population was
not allowed to decrease with time (within the measured errors).  These
fits are turned into SFHs of star formation rate and metallicity as a
function of time. 

To assess the uncertainties of our fits, we then ran Monte Carlo
fitting tests. These tests assess 2 types of errors: random errors due
to Poisson sampling and errors in photometry, and systematic errors
due to deficiencies in the stellar evolution models as well as any
offset in distance, reddening, and/or zero-points.  The Poisson errors
are accounted for by resampling the convolved best-fit model 100
times. Then, when fitting each of these realizations, the systematic
errors are accounted for by introducing small random shifts in the
bolometric magnitudes and effective temperatures of the models.  These
shifts are introduced at the level of the differences between models
in the literature, and therefore serve as a proxy of the effects of
our choice of stellar evolution models.

Our final uncertainties were calculated as the 68\% confidence
intervals of the results of all of our Monte Carlo test fits.  These
total uncertainties are used as the error bars in all subsequent plots
and analysis.

\begin{figure}
\centerline{\psfig{file=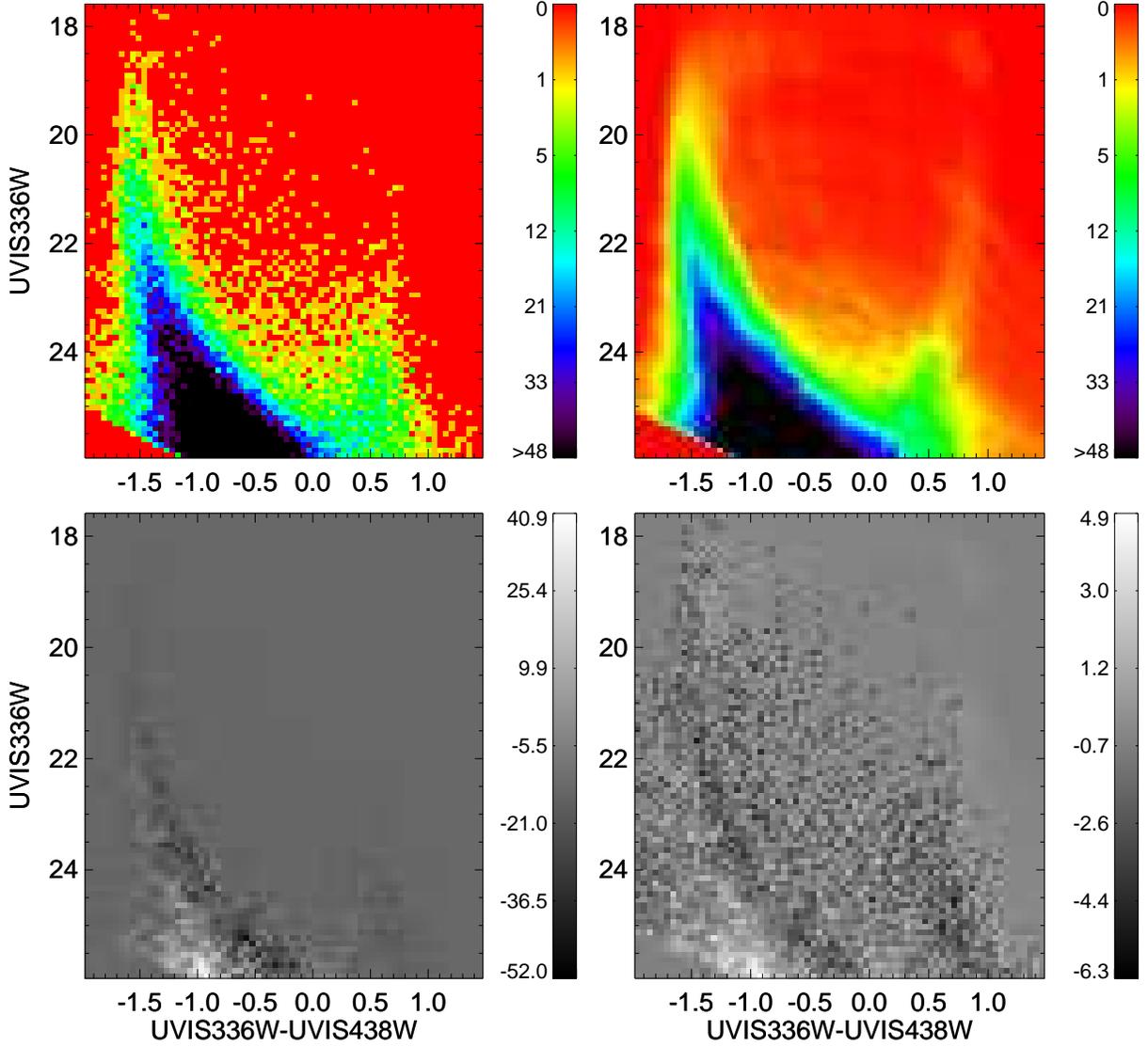,width=6.5in,angle=0}}
\caption{The observed (upper-left), best-fit model (upper-right),
difference (lower-left), and difference normalized by uncertainty
(lower-right) CMDs of our fit to the UVIS data of the inner region of
NGC~4214.  Scale bars are shown to the right of each plot.  The fit is
generally good over the entire CMD assuming foreground extinction, but
also shows a modest amount of internal extinction
(cf.~\S~\ref{bright}) which causes the slight excess in the data
redward of the model main-sequence.}
\label{residuals}
\end{figure}

We can also assess the degree of systematic errors using plots of
residuals between the observed CMD and the model fit.  An example of
the residuals of our fits is shown in Figure~\ref{residuals}.  Our
fits assume the Galactic foreground extinction over the field.
Therefore, the low-level residuals over the entire CMD for these fits
suggests that our photometry is dominated by stars in front of any
significant internal dust layer in the NGC 4214 disk, with a modest
amount of internal extinction (cf.~\S~\ref{bright}) causing the slight
excess in the data redward of the model main-sequence.  Furthermore,
the lack of any strong features in the fit residuals suggests that the
distribution of stellar luminosities is well-fitted assuming a
Salpeter IMF, once the effects of the age distribution are modeled.
These fits show that the main sequence luminosity profile interpreted
by \citet{ubeda2007a,ubeda2007b} as suggesting variations in the IMF,
is also consistent with a standard IMF with a varying star formation
history; thus IMF variations and SFH are somewhat degenerate in such a
complex region.  Our uncertainties come from measurements assuming a
constant IMF.  Therefore, our uncertainties would increase if
variations in the IMF were allowed.

We used a very fine grid of model stellar isochrones (0.05 dex in age
and 0.1 dex in [Fe/H]) to provide the best possible fit to our data.
Because the grid is very fine, pushing star formation between adjacent
time bins has little effect on the quality of the resulting fit.  The
degree to which this degeneracy is true depends on the quality of the
data being fitted, which is characterized by our Monte Carlo tests.

The uncertainties in the cumulative age distribution cannot be
improved by further binning of the data in time.  Therefore, we plot
the full time resolution of our fit to the cumulative age
distribution, showing the uncertainties from our Monte Carlo tests,
which include the effects of pushing star formation between adjacent
time bins.

On the other hand, within a given time bin, the uncertainty on the
measured mean SFR is related to the sensitivity of the CMD to stars in
that age bin.  As the size of the bin increases, more locations of the
CMD will be affected by the bin.  Thus, increasing the bin size
increases the sensitivity of the CMD fit to that time bin, thereby
reducing the uncertainty on the mean rate within that time bin.  It is
therefore possible to reduce the uncertainties on the measured SFR by
increasing the length of the time bins.  

To determine optimal time bins for our differential SFH, we apply the
uncertainties in our cumulative age distribution, as determined from
our Monte Carlo tests.  Specifically, we define bins for which the
cumulative fraction of stars increased with statistical significance.
Specifically, the bins are defined so that at each bin boundary, the
1$\sigma$ upper-limit of the cumulative fraction of stars formed in
the older bin must be less than the 1$\sigma$ lower-limit of the
younger bin.  Thus, each bin contains enough signal in the CMD that if
the bin were removed, the fit to the data would be significantly
degraded. We note that for the optical and UV fits, we adopted the
time bins measured from the IR fits because the optical and UV data
were not sensitive enough to the old stellar populations to provide
cumulative distributions.

\section{Results}

\subsection{Star Formation History}

\begin{figure}
\centerline{\psfig{file=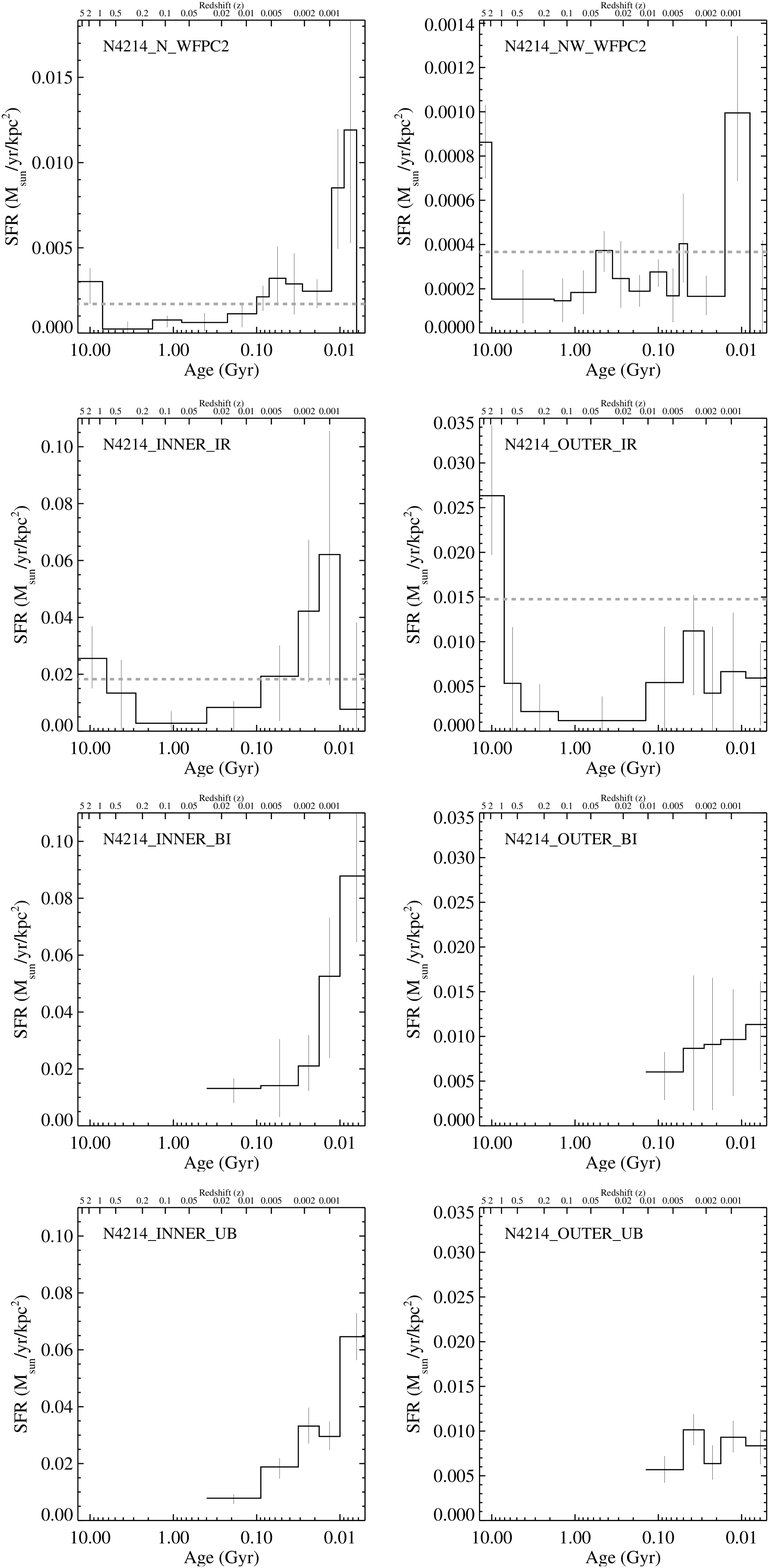,height=8.5in,angle=0}}
\caption{The SFHs of our fields. {\it Upper left:} The SFH of the
WFPC2 field to the north of the center of the galaxy. {\it Upper
right:} The SFH of the deep WFPC2 field to the northwest of the center
of the galaxy.  {\it Second row left:} The SFH of the inner region of
the WFC3/IR field. {\it Second row Right:} The SFH of the WFC3/IR
field excluding the inner region.  {\it Third row left:} The SFH of
the inner region of the WFC3/UVIS field from the F438W and F814W
photometry. {\it Third row right:} The SFH of the WFC3/UVIS field
excluding the inner region from the F438W and F814W photometry.  The
SFH for times prior to $\sim$200 Myr ago is unconstrained by these
data.{\it Bottom left:} The SFH of the inner region of the WFC3/UVIS
field from the F336W and F438W photometry. {\it Bottom Right:} The SFH
of the WFC3/UVIS field excluding the inner region from the F336W and
F438W photometry. The SFH for times prior to $\sim$200 Myr ago is
unconstrained by the UVIS data.}
\label{sfhs}
\end{figure}

\begin{figure}
\centerline{\psfig{file=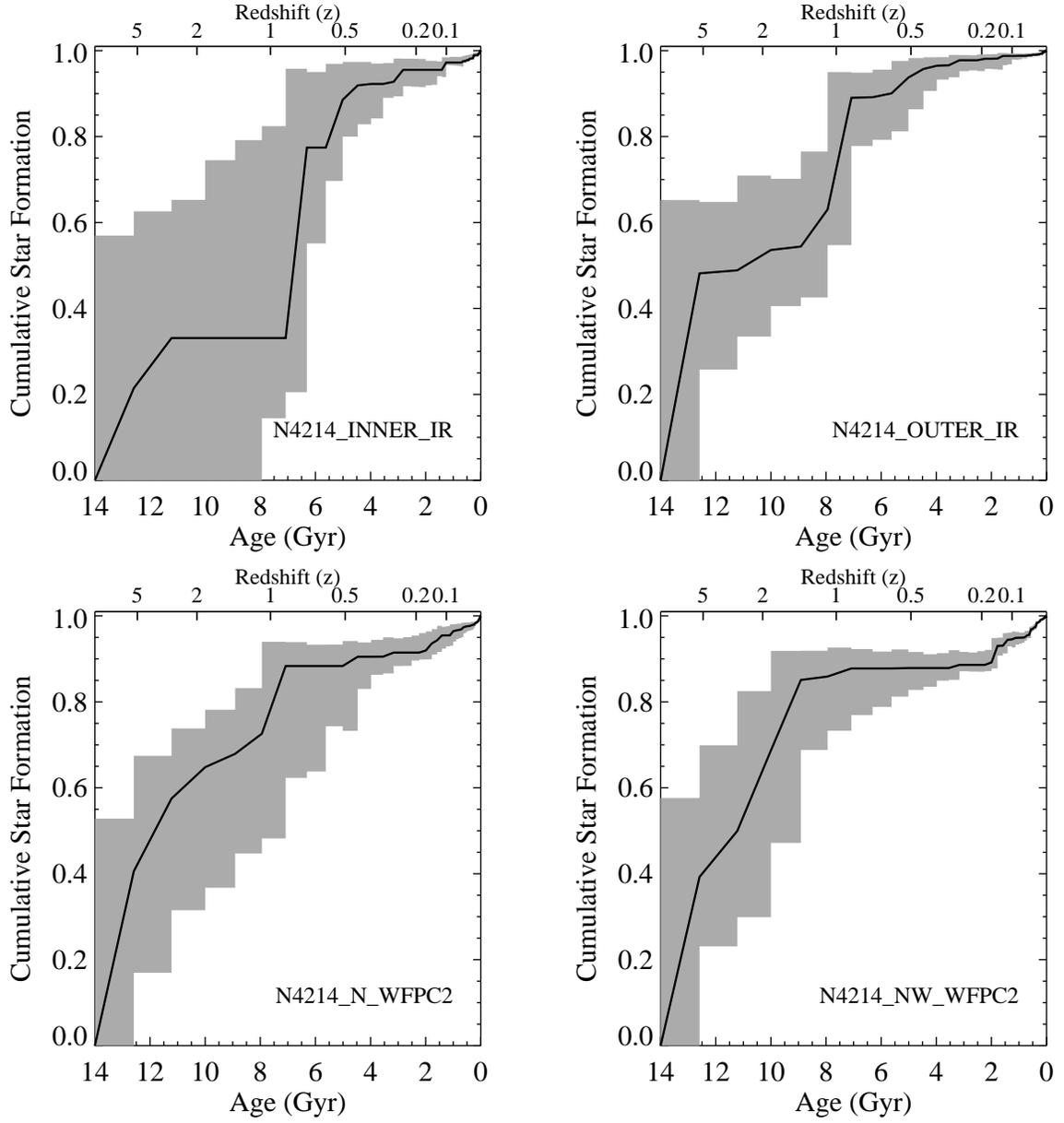,width=6.5in,angle=0}}
\caption{The cumulative formation of stellar mass calculated from the
SFHs of our fields.  The central WFC3 field has been broken up into an
inner region and an outer region, shown in the top two panels. The
UVIS data did not constrain the cumulative formation because it
contained no significant information concerning stars older than 150
Myr.}
\label{cum}
\end{figure}

\begin{figure}
\centerline{\psfig{file=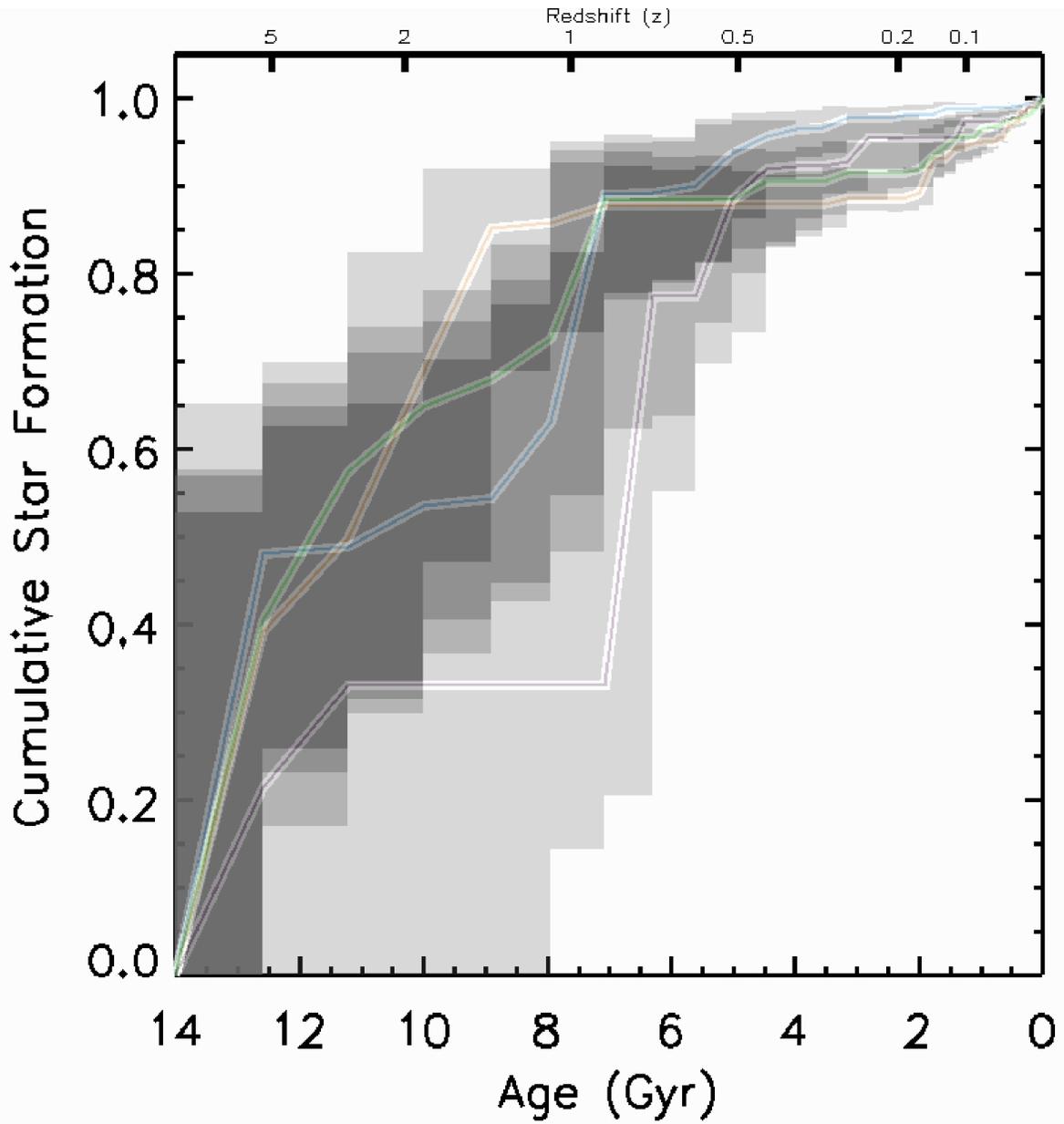,width=6.5in,angle=0}}
\caption{All cumulative formation histories overplotted.  Darker
  shades of gray represent overlaps in the cumulative SFHs at
  different radii.  All of our measurements
are consistent with 75\% of the stellar mass forming by z$\sim$1.  The best-fit solutions are shown with colored lines (orange=N4214\_NW\_WFPC2, green=N4214\_N\_WFPC2, blue=N4214\_OUTER\_IR, purple=N4214\_INNER\_IR).}
\label{allcum}
\end{figure}

The resulting SFHs for our CMD fits to the 4 regions are shown in
Figure~\ref{sfhs}.  The UVIS data did not provide sufficient depth to
probe ages greater than $\sim$200 Myr, and thus no points are plotted
for older ages.  Cumulative plots of the stellar mass formed are shown
in Figure~\ref{cum}, and the results for all regions are overplotted
in Figure~\ref{allcum}. Since the UVIS data were not sensitive to old
populations, we could not generate cumulative SFHs from the UVIS data.

The SFHs show that in all of the areas observed, the majority of the
stellar population is old.  In the central portion of the galaxy, the
relatively shallow data are only able to constrain the population only
weakly, putting a lower limit of $>$80\% of the stellar mass having
ages $>$4~Gyr.  In the outermost (and deepest) field, the constraint
is tighter, with $>$74\% of the stellar mass having ages $>$8~Gyr.
Taking uncertainties into account, all of our data are consistent with
this constraint from the deepest photometry.

\begin{figure}
\centerline{\psfig{file=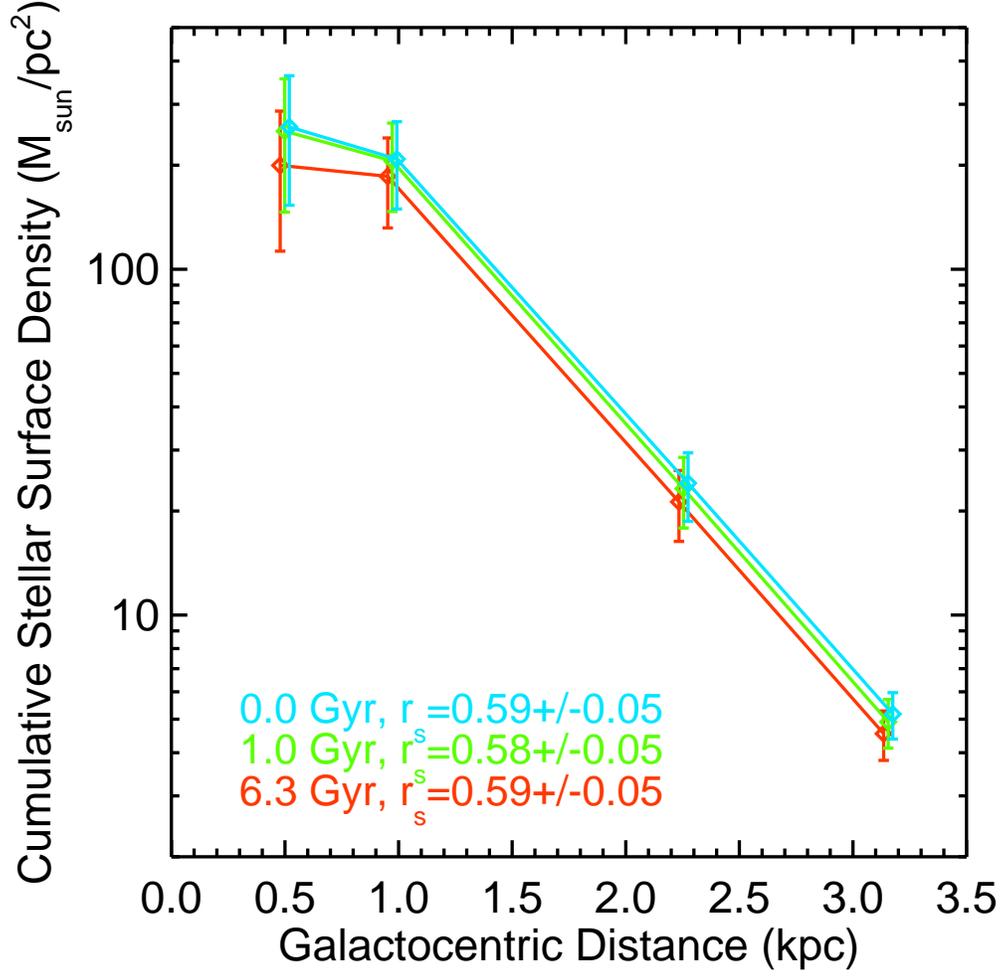,width=5.5in,angle=0}}
\caption{The cumulative stellar mass density profile of NGC~4214 at 3
epochs, as calculated from our SFH measurements.  Changes in the IMF
and/or mass cutoff could shift all of the surface density data up or
down. The scale-lengths are unaffected by these IMF issues.}
\label{dens_vs_r}
\end{figure}

The consistency of all of our data, from the optical to the IR, from
the center to the outskirts of the disk, with more than three quarters
of the stellar mass being very old is surprising.  This galaxy is
well-known for being a young starburst galaxy, being known as a
``Wolf-Rayet'' galaxy \citep{sargent1991}.  While the current star
burst is already known to be responsible for only a few percent of the
stellar mass \citep{huchra1983}, it is still surprising that such a
small percentage of the population is young ($<$1\% in the past 50
Myr, $<$4\% in the past Gyr), and such a high percentage is very old.
Overall, the galaxy has evolved very little since $z\,\sim\,1$, as shown
in our plot of the surface density profile over the past 6.3 Gyr
(Figure~\ref{dens_vs_r}).

Our measurements also provide some indication of the metallicity of
the stellar populations in NGC~4214.  Overall, the metallicity of the
populations fell in the range of -1.6$<$[M/H]$<$0, with the old
population being -1.6$<$[M/H]$<$-0.6 in the IR data and
-1.4$<$[M/H]$<$-0.6 in the WFPC2 data, and the young population being
-0.5$<$[M/H]$<$0 in the UVIS data.  The WFPC2 data had very little
sensitivity to metallicity for the young population, as MS stars of
different metallicities have very similar colors in F606W-F814W.


\subsection{The Young Stellar Population}\label{bright}

The old population likely dominates at all radii, suggesting that this
population is well-mixed and NGC~4214 is old.  However, the young
stellar populations are clearly not well-mixed, as their density
clearly varies significantly with position in the galaxy. This
difference can be seen by looking at the strong core He-burning
sequence extending vertically from the tip of the RGB in the central
WFC3/IR CMD and northern WFPC2 CMD but relatively absent in the
northwest WFPC2 CMD.  The central starburst is evident, even when the
population is probed only in the IR, where massive young stars are
faint, although we note that the very youngest population ($<$10~Myr)
in the innermost region is not reliably measured in the IR
(Figure~\ref{sfhs}).  The SFH of the innermost region of the galaxy is
the only one with a sustained recent ($\lap$30~Myr) star formation
rate that is higher than any rate in previous epochs.  This result
does not mean that the rate is higher than it has ever been in the
past, since our time resolution would not allow us to detect short,
strong bursts at ancient times, but the result does show that very
recent star formation near the galaxy center is at least a factor of 2
higher than the average rate over the galaxy's history.  Just outside
of this central starburst, the star formation rate is slightly less
than the overall mean rate but higher than the mean rate over the past
several Gyr.

Outside of the central WFC3 field, the knots of star formation in the
northern WFPC2 field are easily seen in the CMD.  Recent star
formation produces both the strong red He-burning sequence extending
brightward from the RGB and the well-populated upper MS.  The SFH
shows a strong increase in star formation rate starting 100~Myr ago
with a prominent peak at very recent times (5--12 Myr ago), providing
some indication of the age of these features.  Prior to 100~Myr ago,
our analysis shows that these regions of NGC~4214 were relatively
quiescent, perhaps bearing resemblance to an early-type dwarf.

\begin{figure}
\centerline{\psfig{file=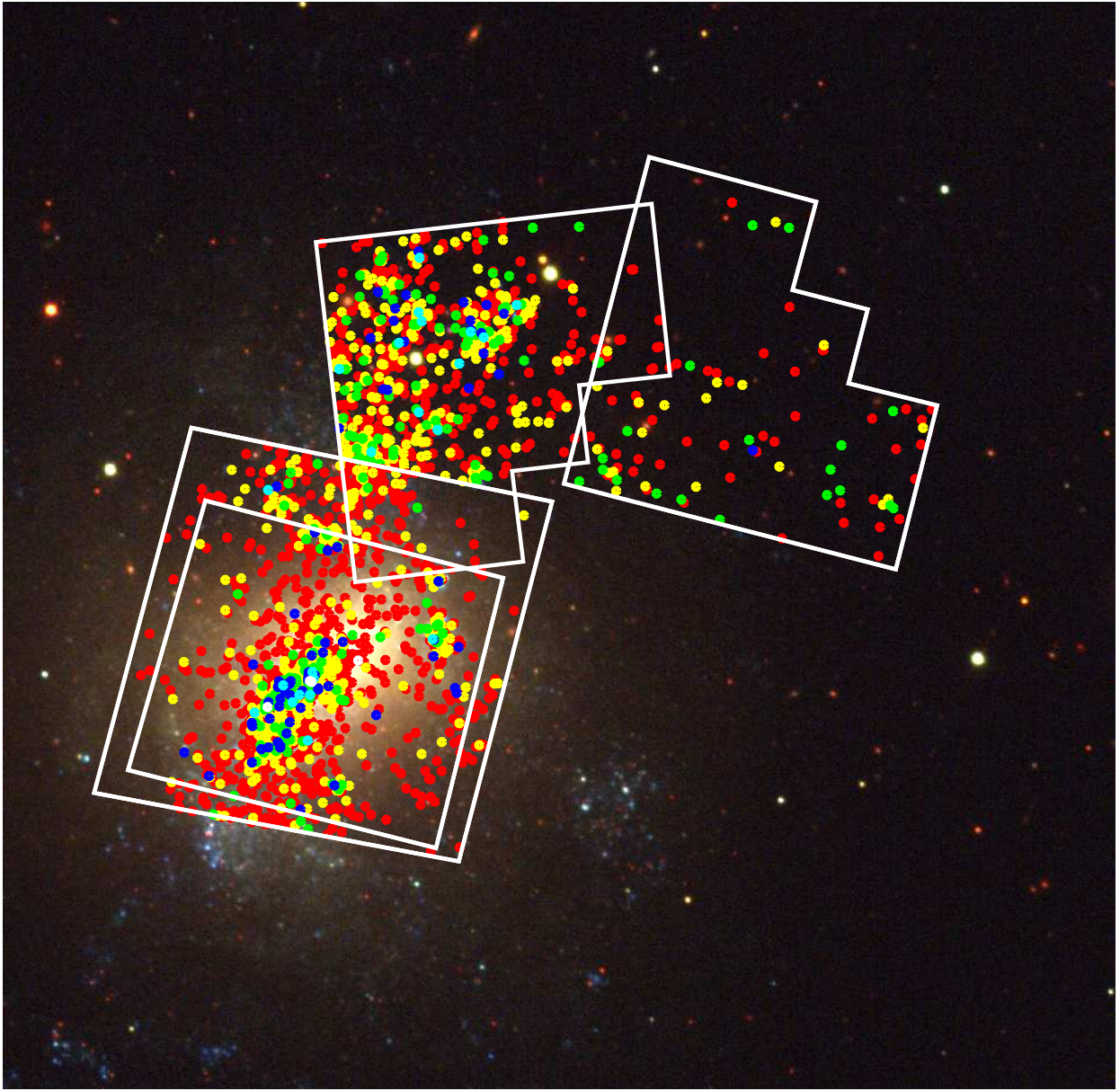,width=6.5in,angle=0}}
\caption{The locations of the massive ($>$12~M$_{\odot}$) young stars
in our images.  Bluer colors indicate more massive (luminous) stars
(red = 12--22~M$_{\odot}$; yellow = 23--34~M$_{\odot}$; green =
35--40; blue = 41--48~M$_{\odot}$; cyan = 49--56~M$_{\odot}$; white =
$\sim$56~M$_{\odot}$).}
\label{young_stars}
\end{figure}

In Figure~\ref{young_stars} we plot the spatial distribution of young
stars as a function of brightness.  The brightest (youngest) stars
appear to be more tightly clustered than the fainter stars.  A
2-dimensional K-S test yields only a 4\% probability that the
brightest MS stars (cyan and white diamonds in
Figure~\ref{young_stars}) and the faint MS stars (red diamonds in
Figure~\ref{young_stars}) are drawn from the same spatial
distribution.  Thus stars have likely migrated significantly from
their birth locations on timescales of just $\sim$12~Myr (the
main-sequence lifetime of a 12~M$_{\odot}$ star).

\begin{figure}
\centerline{\psfig{file=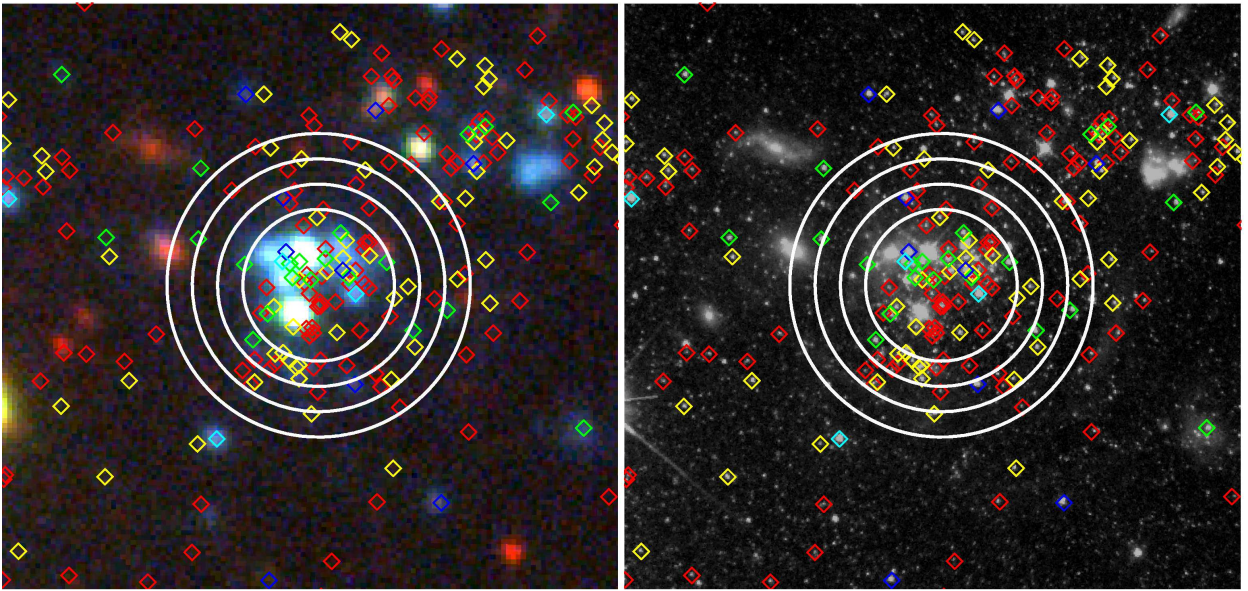,width=6.5in,angle=0}}
\caption{The locations of the massive ($>$12~M$_{\odot}$) young stars
in the vicinity of a star forming knot in our WFPC2 field north of the
galaxy plotted on the SDSS color image ({\it left}) and our F606W
WFPC2 image ({\it right}).  Annuli mark our radial bins to search for
a correlation between age and distance from knot center.  Colors are
the same as in Figure~\ref{young_stars}.}
\label{sfreg}
\end{figure}

To gain a sense of the velocities with which stars are spreading from
their birth regions, we looked for a good example region to study in
detail.  One of the knots of star formation (RA = 183.8884, DEC =
36.3683, see Figure~\ref{sfreg}) was isolated enough to look for a
correlation between the magnitude of the brightest MS stars and the
distance from the center of the knot.  Taking the brightest MS star in
annuli of 0--6$''$, 6--8$''$, 8--10$''$, and 10--12$''$ (see
Figure~\ref{sfreg}), there is a weak correlation between the magnitude
of the brightest MS star and distance from the central knot.  Each 75
pc farther out, the brightest MS star is $\sim$1 magnitude fainter. If
we assume the brightest MS star is a proxy for the MS turnoff
magnitude, at these bright magnitudes on the upper main sequence 1
magnitude corresponds roughly to 10~Myr of age.  If we further assume
that the weak correlation is real and that it is due to the diffusion
of stars produced in the center of the knot, the slope of the
correlation is consistent with a diffusion speed of $\sim$8 pc
Myr$^{-1}$ ($\sim$8 km s$^{-1}$).  This value is similar to the
velocity dispersion of B stars in the Milky Way disk
\citep{dehnen1998} and the diffusion speed found for an outer spiral
arm in M81 \citep{williams2009a}.

The brightest MS stars in our catalog provide diagnostic tests for
massive stellar evolution models. HST UV spectroscopy has been
measured for the central starburst, yielding an age of 4--5 Myr
\citep{leitherer1996}.  Our photometry reveals point sources with
F336W ($U$-band equivalent) magnitudes of $\sim$17.6, which is
$M_{F336W}$=-9.9 at the distance and extinction of NGC~4214.  We show
our F336W-F438W CMD in Figure~\ref{cmds} along with the isochrone for
an age of 4~Myr and a metallicity of [M/H]=-0.4 shifted to the
distance and foreground extinction of NGC~4214.  Assuming these
objects are not unresolved compact clusters, these bright stars are
consistent with the brightest blue portion of this isochrone, which
represents model stars with masses of 52--56~M$_{\odot}$.  The low
foreground extinction ($A_V = 0.07$) and good agreement between the
data and model isochrone suggests that these stars are not strongly
affected by extinction internal to NGC~4214. NGC~4214 is known to
contain just a modest amount of internal extinction \citep[internal
{\sl GALEX} FUV extinction $\sim$0.58 mag,][]{lee2009}, and these
particular UV-bright stars are apparently between us and any of
significant dust within NGC~4214, making them easily detected and
measured in the UV.

Finally, we estimated the approximate total SFR for the galaxy over
the past $\sim$100~Myr, as it has been relatively high during most of
this period. If we add together the SFRs from our 4 regions, we obtain
a total rate of $\sim$0.1~M$_{\odot}$~yr$^{-1}$, $\sim$80\% of which
is in the central UVIS field.  This rate is consistent with the SFR
recently measured for NGC~4214 from {\sl HST/WFPC2} photometry by
\citet{mcquinn2010a,mcquinn2010b}.

\subsection{Dominant Galaxies in their Environment}

Interestingly, the overall population is similar to that of the very
different nearby dwarf galaxy NGC~404, which is a somewhat more
massive ($\sim$4.5$\times$10$^9$ M$_{\odot}$;
\citealp{karachentsev2004}) S0 galaxy with very little recent star
formation. The ANGST data for that galaxy show that outside the inner
regions $\sim$90\% of the stellar population is $>$8 Gyr old and
$\sim$75\% is $>$10~Gyr old \citep{williams2010}.  While the overall
stellar populations of NGC~404 are older than those of NGC4214, the
difference is not overwhelming.  In fact, by 4~Gyr ago, both galaxies
had formed 90\% or more of their stellar mass.  Apparently, the vast
difference in morphology between these 2 galaxies is mainly due to
just a small percentage of the stellar mass.  Our results show that
$\sim$1\% of the stellar mass in NGC~4214 formed in the past
$\sim$100~Myr.  The percentage of stellar mass formed in the past
100~Myr is much smaller in NGC~404, only $<$0.1\% outside of the
galaxy center.  However, there is evidence for a ~1 Gyr old starburst
in the center of NGC 404 \citep{seth2010}; about 1 Gyr ago NGC 404 may
have more closely resembled NGC 4214's present-day appearance.  While
this difference represents a small percentage of the total stellar
populations of the galaxies, it results in their vastly different
morphological classifications.

Both NGC~4214 and NGC~404 appear to have older median ages than some
more massive disks, such as M33 \citep{williams2009b}.  Perhaps their
old stellar ages are somehow due to their relationship with their
environments.  Both galaxies are the most massive in their local
environment.  Perhaps it is common for the dominant galaxy by mass
within any group of galaxies (where NGC~404 is isolated, making its
own ``group'') to be dominated by ancient stars.  If this effect is
indeed common, it may indicate that the most massive galaxy in a group
dominates gas accretion in the early stages of evolution, winning the
local ``downsizing'' battle to be the dominant star forming object
during the epoch of formation.

\section{Conclusions}

We have analyzed deep HST/WFPC2 photometry of the NGC~4214 disk and
HST/WFC3 photometry of the central portion of NGC~4214.  Full CMD
modeling of the photometry shows that the stellar populations
throughout the disk are old, with $\sim$75\% of the stellar mass older
than $\sim$8~Gyr.  This result shows that overall the stellar
populations of NGC~4214 are similar to those of the S0 galaxy NGC~404,
other than the youngest 1\% of the stellar mass, which formed in the
past 100~Myr in NGC~4214.  The similarity suggests that a few hundred
Myr ago, NGC~4214 may have looked very similar to NGC~404 today,
though NGC~4214 may have been more gas-rich.  Alternatively, NGC~4214
may have recently acquired gas from a merging satellite, such as the
merger suggested by \citet{fanelli1997}.  NGC~4214 currently has more
than 10 nearby satellites \citep[e.g.][Figure 2]{dalcanton2009};
therefore such an alternative is a strong possibility.  Finally, we
presented an argument that because both galaxies are the dominant
members of their local environments, they were in favorable positions
to form such a high percentage of their stars so early.
 
Support for this work was provided by NASA through grants GO-10915,
GO-11719, and GO-11986 from the Space Telescope Science Institute,
which is operated by the Association of Universities for Research in
Astronomy, Incorporated, under NASA contract NAS5-26555.


\clearpage

\begin{deluxetable}{lcccccccc}
\tablecaption{Magnitude Limits of the Observed Regions}
\tablehead{
\colhead{{\footnotesize Region}} &
\colhead{{\footnotesize $F336W_{50}$\tablenotemark{a}}}  &
\colhead{{\footnotesize $F438W_{50}$\tablenotemark{b}}}  &
\colhead{{\footnotesize $F606W_{50}$\tablenotemark{c}}} &
\colhead{{\footnotesize $F814W_{50}$\tablenotemark{d}}} &
\colhead{{\footnotesize $F110W_{50}$\tablenotemark{e}}} &
\colhead{{\footnotesize $F160W_{50}$\tablenotemark{f}}}  &
\colhead{{\footnotesize GCD\tablenotemark{g}}} 
}
\startdata
{\footnotesize UV INNER}  & 26.0 &   27.1  & \nodata &  \nodata &
\nodata & \nodata & 0.34 \\
{\footnotesize UV OUTER}  & 26.1 &   27.2  & \nodata &   \nodata &
\nodata & \nodata & 1.15\\
{\footnotesize IR INNER}   & \nodata &   \nodata & \nodata &  \nodata
& 23.5 & 22.7 & 0.52\\
{\footnotesize IR OUTER}  & \nodata &   \nodata  & \nodata &   \nodata
& 24.0 & 23.2 & 0.99\\
{\footnotesize BI INNER}  & \nodata &   26.9  & \nodata &  26.0 &
\nodata & \nodata & 0.48 \\
{\footnotesize BI OUTER}  & \nodata &   27.1  & \nodata &  26.2 &
\nodata & \nodata & 1.10\\
{\footnotesize WFPC2 N}  & \nodata &   \nodata  & 26.7 & 26.1 &
\nodata &   \nodata & 2.27\\
{\footnotesize WFPC2 NW}  & \nodata &   \nodata  & 27.9 & 27.1 &
\nodata &   \nodata & 3.17\\
\enddata
\tablenotetext{a}{The 50\% completeness limit of the F336W data.}
\tablenotetext{b}{The 50\% completeness limit of the F438W data.}
\tablenotetext{c}{The 50\% completeness limit of the F606W data.}
\tablenotetext{d}{The 50\% completeness limit of the F814W data.}
\tablenotetext{e}{The 50\% completeness limit of the F110W data.}
\tablenotetext{f}{The 50\% completeness limit of the F160W data.}
\tablenotetext{g}{The median deprojected galactocentric distance of the stars in the region in kpc.}
\label{table}
\end{deluxetable}

\end{document}